\documentclass[aps,prl,twocolumn,superscriptaddress,floatfix]{revtex4} 
\usepackage{xcolor}
\usepackage{graphicx}%
\usepackage{bm}%
\usepackage{latexsym}
\usepackage{amsmath}
\usepackage{extarrows}
\usepackage{natbib}
\usepackage{braket}
\usepackage{url}
\usepackage{tikz}

\begin{document}

	\title{Dynamical local connector approximation for electron addition and removal spectra}
	
	\newcommand{\lsi}{Laboratoire des Solides Irradi\'es, \'Ecole Polytechnique, CNRS, CEA,  Universit\'e Paris-Saclay, F-91128 Palaiseau, France}
	\newcommand{\etsf}{European Theoretical Spectroscopy Facility (ETSF)}
	\newcommand{\soleil}{Synchrotron SOLEIL, L'Orme des Merisiers, Saint-Aubin, BP 48, F-91192 Gif-sur-Yvette, France}
	
	\author{Marco Vanzini}
	\email[]{marco.vanzini@polytechnique.edu}
	\affiliation{\lsi}
	\affiliation{\etsf}
	
	\author{Lucia Reining}
	\affiliation{\lsi}
	\affiliation{\etsf}

	\author{Matteo Gatti} 
	\affiliation{\lsi}
	\affiliation{\etsf}
	\affiliation{\soleil}

	\date{August 4, 2007}
	
	\begin{abstract}
Realistic calculations of electron addition and removal spectra rely most often on Green's functions and complex, non-local self-energies. We introduce a shortcut to obtain the spectral function directly from a local and frequency-dependent, yet real, potential. We calculate this potential in the homogeneous electron gas (HEG), and we design a connector  which prescribes the use of the HEG results to calculate spectral functions of real materials. Benchmark results for several solids demonstrate the potential of our approach. 
	\end{abstract}
	
	\maketitle
	
	Photoemission and inverse photoemission experiments are powerful tools to investigate materials \cite{Huefner2003}. They give access to electron removal and addition spectra, 
    described by the spectral function (SF) $A\left(\omega\right)$.
	The SF is an integral over momentum-dependent components 
	$A\left(\boldsymbol{k},\omega\right)$, which can be studied by angle-resolved photoemission experiments \cite{Damascelli2013}. Alternatively, the SF can also be obtained as an integral over the \emph{local}  $A\left(\boldsymbol{r},\omega\right)$, which is measured in scanning tunneling spectroscopy \cite{Binnig1987,Chen1993}. 
	To analyze or predict photoemission spectra is a major challenge, as the electron-electron interaction causes spectra to be quantitatively and even qualitatively different from any independent-particle result \cite{OnReRu}  \footnote{One should also describe the photoemission process itself (e.g. the photon-energy dependence of the spectra), see e.g. Refs. \protect\cite{Huefner2003,Hedin1998}, which is however not the subject of the present work.}.
	
	One of the most widely used theoretical frameworks to describe photoemission  is
	many-body perturbation theory (MBPT) \cite{Fetter,Martin2016}, 
	where the one-body Green's function \footnote{Spin is omitted for simplicity throughout the paper since here we deal only with spin-unpolarized materials. The generalization to spin-polarized situations is possible.} $G\left(\boldsymbol{r},\boldsymbol{r}',\omega\right)$ is the key quantity for electron addition and removal spectroscopies, as  the local SF is related to the diagonal of $G$ through
	\begin{equation}
	A\left(\boldsymbol{r},\omega\right)=\frac{1}{\pi}
|\operatorname{Im} G\left(\boldsymbol{r},\boldsymbol{r},\omega\right)|.
	\label{eq:LSF}
	\end{equation}
	The local SF yields the electronic density through $n\left(\boldsymbol{r}\right)=\int_{-\infty}^{\mu}d\omega A\left(\boldsymbol{r},\omega\right)$ (with $\mu$ the Fermi energy), which is also important for accessing ground-state properties.
	Much work is therefore devoted to the calculation of $G$.
	The GF is usually obtained from a Dyson equation, where all exchange-correlation (xc) effects are contained in the non-local, frequency dependent and
	non-hermitian self-energy  $\Sigma_{\rm xc}\left(\boldsymbol{r},\boldsymbol{r}',\omega\right)$. 

	Today, well established approximations for the self-energy, such as Hedin's GW approximation \cite{HedinGW}, give access to the SF of a wide range of 
	materials \cite{AryaGunn,Schilfgaarde2006}. Corrections of higher order in the interaction \cite{Martin2016} make the calculations quickly unfeasible for realistic systems. 
    Even on the level of GW, the non-locality of the self-energy renders calculations much slower than, 
	e.g., density-functional theory (DFT) with a local, static and real Kohn-Sham (KS) potential  $v_{\rm KS}(\boldsymbol{r})$ \cite{DFT,KohnSham}. Of course, such a simple potential cannot, as a matter of principle, yield the correct $G$ and indeed, attempts to use the KS band structure for the description of photoemission spectra and band gaps are problematic \cite{SK,gap1,ShamSchluter,OnReRu}. 
	
	However, Eq. \eqref{eq:LSF} shows that the traditional path of MBPT is a detour for the calculation of the local SF: one has to evaluate the whole $G$, and subsequently 
	discard most of the information. This is fundamentally inefficient, and not satisfactory from the point of view of principle. 
	Spectral density functional theory has been proposed \cite{Savrasov} as an in principle exact alternative to calculate the local SF. It makes use of a short-range self-energy, but the latter is still non-hermitian, and no feasible approximations for realistic systems are known.
Therefore, to find a shortcut and obtain the local SF, photoemission spectra and the electronic density, \emph{without} passing through the full $G$ and $\Sigma_{\rm xc}$, remains un unsolved problem. 
	
	Ref. \cite{MatteoPRL} suggested that such a shortcut exists in principle: it was shown that a local and real \emph{spectral potential} (SP), with a frequency-dependent xc contribution $v_{\rm SF}\left(\boldsymbol{r},\omega\right)$, can be constructed from  
 a generalized Sham-Schl\"uter equation  \cite{ShamSchluter}, and that it can be used to calculate the local SF in principle exactly. 
 In \cite{Ferretti} this equation was solved  for $v_{\rm SF}$ in a simple model. 
	However, to the best of our knowledge today no approach is available that would determine $v_{\rm SF}$ for a model or real material without passing through the calculation of the full self-energy. 
	
\begin{figure}[t!]
\includegraphics[width=1.0\columnwidth]{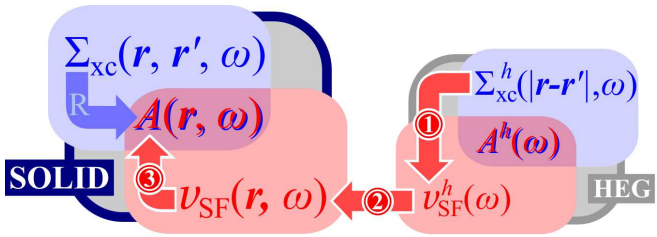}
	\caption{(Color online) Schematic view of the dynamic connector approach: 
    \textcircled{\raisebox{-0.9pt}{1}}
 calculate the SP $v_{\rm SF}^h(\omega)$ in the HEG for a given approximation to $\Sigma_{\rm xc}^h(|\boldsymbol{r}-\boldsymbol{r}'|,\omega)$. The two share the same SF $A^h(\omega)$; \textcircled{\raisebox{-0.9pt}{2}} design the connector and import the SP in the real system (\emph{e.g.}, a solid); \textcircled{\raisebox{-0.9pt}{3}} use the SP $v_{\rm SF}(\mathbf{r},\omega)$ to evaluate the SF $A(\boldsymbol{r},\omega)$. This approach yields in principle the same result as the non-local SE $\Sigma_{\rm xc}(\boldsymbol{r},\boldsymbol{r}',\omega)$ used as reference, as indicated by the letter R.}
	\label{fig:schema}
	\end{figure}

	
	The aim of the present work is to design a way to obtain $v_{\rm SF}$, and 
	to demonstrate that this leads to a feasible and powerful method to calculate photoemission spectra. We support these claims with calculations for four very different materials. 
	Our strategy is summarized in Fig. \ref{fig:schema}. It is inspired by the way in which density functionals such as the local-density approximation  (LDA) overcome the absence of a diagrammatic method: first, the quantity of interest [in KS, the density] is obtained from an auxiliary system described by a fictitious potential [in KS, the KS potential including the xc contribution $v_{\rm xc}(\boldsymbol{r})$]. The auxiliary 
	xc potential is calculated for a model system [the homogeneous electron gas (HEG) in the case of the LDA], by an advanced, more expensive, method, such as Quantum Monte Carlo \cite{CepAld}. The auxiliary potential of the model system is then used to simulate the auxiliary potential of the real system by using some approximate prescription, such as the LDA. 
	
	In our case, the quantity of interest is $A\left (\boldsymbol{r},\omega\right)$, and the auxiliary system is described by the SP $v_{\rm SF}\left(\boldsymbol{r},\omega\right)$. We propose (1)
		to calculate and tabulate the SP $v_{\rm SF}^h\left(\omega\right)$ of the HEG for a series of densities: this calculation has to be done \emph{only once and forever}, similarly to the Monte Carlo calculations of Ceperley and Alder \cite{CepAld};  
		(2) to design a \emph{connector}, \emph{i.e.}, a prescription of how to use the resulting table in order to construct the SP $v_{\rm SF}\left(\boldsymbol{r},\omega\right)$ of the real material; 
		(3) for the connector proposed in the present work: to use the resulting SP in a particular way, as explained below, in order to obtain $A\left(\boldsymbol{r},\omega\right)$. 

\begin{figure}[t!]
		\includegraphics[width=1.0\columnwidth]{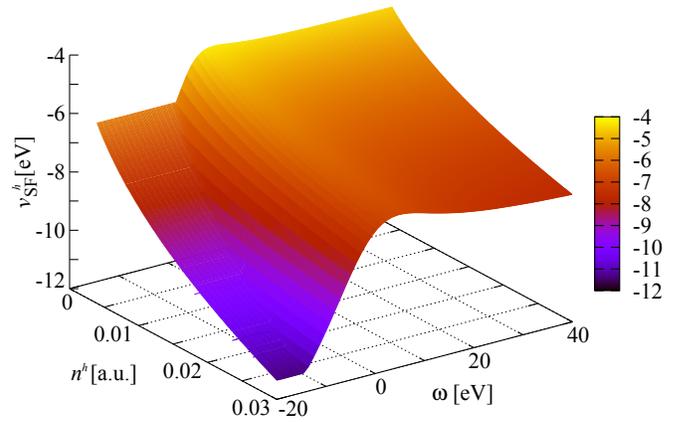}
		\caption{(Color online) HSE06 spectral potential $v^h_{SF}\left(\omega\right)$ (in eV) in the HEG as a function of frequency (in eV), for different values of densities (in $a_0^{-3}$), ranging from $n^h=3.93\cdot10^{-3}a_0^{-3}$ ($r_s=3.93a_0$, corresponding to sodium) to $n^h=3.26\cdot10^{-2}a_0^{-3}$ ($r_s=1.94a_0$, argon).}
		\label{fig:potential}
	\end{figure}

	For (1), we start by evaluating the full self-energy $\Sigma^h_{\rm xc}$ in the HEG. For the purpose of the demonstrations in this Letter we use a realistic non-local approximate self-energy, namely a static range-separated hybrid, the HSE06 \cite{HSE06}
\footnote{The explicit form for the exchange-correlation SE is $\Sigma_{\rm xc}^{\rm HSE06}\left(\boldsymbol{r},\boldsymbol{r}'\right)=v_{\rm xc}\left(\boldsymbol{r}\right)+ 0.25 \left[\Sigma^{\rm SR}_{\rm x}\left(\boldsymbol{r},\boldsymbol{r}'\right)-v_{\rm x}^{\rm SR}\left(\boldsymbol{r}\right)\right]$, where in the Fock SE $\Sigma^{\rm SR}_{\rm x}$ and in the local exchange-only $v_{\rm x}^{\rm SR}$ the Coulomb interaction is screened by the complementary error function giving a short-range (SR) interaction. Note that at variance with Ref. \protect\onlinecite{HSE06} here we have used the LDA for the local potentials.}. 
    This relatively simple approximation allows us to highlight the non-trivial task at this stage, namely, the conversion of non-locality into pure frequency dependence in the SP \cite{MatteoPRL}.
In the HEG the SP is obtained as a compact expression in terms of  $\Sigma^h_{\rm xc}$. 
Details can be found in the supplemental material \cite{suppmat}.
We stress again that in our scheme for a given approximation of the self-energy the SP has to be calculated \emph{only once and forever} 	\footnote{Ultimately, one will of course do the best possible calculation for $\Sigma^h_{\rm xc}$ at this stage. We also note that correlation contributions in the self-energy beyond HSE06 become more local in space \protect\cite{Martin2016}, making the conversion of non-locality into frequency dependence simpler.}.
Fig. \ref{fig:potential} shows the strong frequency-dependence of the SP  obtained from the non-local HSE06 self-energy, despite the fact that the HSE06 self-energy is static. 
We have  tabulated this HSE06-derived SP and made it freely available \footnote{The database of potentials with entries $\left(n,\omega,v^h_{\rm SF}\left(\omega\right),\Delta^h\left(\omega\right)\right)$ can be downloaded at http://etsf.polytechnique.fr/research/connector/dynLCA.}. 
 

In order to use this table, we move to (2). The design of a \emph{connector} in this step is the most difficult part of our work. 
	The simplest idea would be to use for 
	$v_{\rm SF}\left(\boldsymbol{r},\omega\right)$ at each point $\boldsymbol{r}$ the HEG result evaluated for the local density, as in LDA, but this choice is not flexible enough \cite{suppmat}.
	In principle one should use the HEG SPs calculated with a different density for each point in space and for each frequency, 
	$v_{\rm SF}\left(\boldsymbol{r},\omega\right)=v^h_{\rm SF| \it{n^h}=\mathcal{F}\left(\boldsymbol{r},\omega,[\it{n}]\right)}\left(\omega\right)$. If $v_{\rm SF}^h$ spans the range of values taken by $v_{\rm SF}$, there should be a function $\mathcal{F}$ which makes this connector exact. However, it might be exceedingly complicated. Here we take a different route: we modify the HEG SP based on physical insight, bringing it closer to the SP of the real system, such that the simple 
	LDA
	connector $n^h=n\left(\boldsymbol{r}\right )$ is sufficient. As we will show below, using besides the average density only ingredients based on local quantities, i.e., making a \textit{dynamical local connector approximation}, yields already very promising results. 
	
The first intuitive modification of $v_{\rm SF}^h$ towards the real system is to align the total potentials, namely 
$v_{\rm SF}\left(\boldsymbol{r},\omega\right)=v^h_{\rm SF}\left(\omega\right)-v_{\rm e}\left(\boldsymbol{r}\right)-v_{\rm H}\left(\boldsymbol{r}\right)$. Here, $v_{\rm e}$ and $v_{\rm H}$ are the external and the Hartree potential in the real system; they sum to zero in the HEG. Next, it is reasonable to suppose that
the energy of the HEG and, locally, of the real system, should be aligned \emph{in the spectral function}. 
	This is trivial in the HEG, where a local potential is just a number $c$ which simply
    shifts  
the frequency scale. 
	In the real system, shifting the potential by $c\left(\boldsymbol{r}\right)$ or shifting the frequency argument in the spectral function is not equivalent, and we have to specify $c\left(\boldsymbol{r}\right)$ in the connector \cite{suppmat}:
	\begin{equation}
	v_{\rm SF}\left(\boldsymbol{r},\omega\right)=v^h_{\rm SF}\bigl(\omega-c\left(\boldsymbol{r}\right)\bigr)+c\left(\boldsymbol{r}\right)-v_{\rm e}\left(\boldsymbol{r}\right)-v_{\rm H}\left(\boldsymbol{r}\right).
	\end{equation}
    Here we use the KS potentials for the alignment,
	\begin{equation}
	c\left(\boldsymbol{r}\right)=v_{\rm KS}(\boldsymbol{r})-v^h_{\rm KS}\left[\bar n\right],
	\label{eq:connector}
	\end{equation}
    where $\bar{n}$ is the average density    \footnote{This choice  reduces to $v_{\rm e}(\boldsymbol{r})+v_{\rm H}(\boldsymbol{r})\sim \mu-\mu^h$ for slowly varying density. The SP simplifies and becomes $v_{\rm SF}(\boldsymbol{r},\omega)\approx v^h_{\rm SF}\bigl(\omega-\mu+\mu^h\bigr)$, 
		namely the HEG potential with a rigid alignment of the Fermi energies, as proposed in \cite{SK} for the local density approximation to the SE. The performance of this simpler connector is discussed in \cite{suppmat}.}.
	Finally, we rescale frequencies by the plasmon energies, which set the characteristic energy scales \cite{Cappellini1993}. With this,
	\begin{multline}
	v_{\rm SF}(\boldsymbol{r},\omega)=\\v^h_{\rm SF | \it{n^h=n}(\boldsymbol{r})}\left[\frac{\omega_P\left(n\left(\boldsymbol{r}\right)\right)}{\omega_P\left(\bar n\right)}\Bigl(\omega-v_{\rm KS}(\boldsymbol{r}) 
	+v^h_{\rm KS}\left[\bar n\right]\Bigr)\right] \\
	+v_{\rm xc}(\boldsymbol{r})-v^h_{\rm xc}\left[\bar n\right].
	\label{eq:pot+++}
	\end{multline}
	This \emph{dynamical local connector approximation} (dynLCA)  yields a spectral potential where all ingredients are explicit density functionals, calculated once forever in the HEG, or results of a KS calculation.  
	For the demonstration in this work we calculate the spectral function in first order perturbation theory. Therefore, the matrix elements  $\left<\ell\boldsymbol{k}\right|v_{\rm SF}\left(\boldsymbol{r},\omega\right)\left|\ell\boldsymbol{k}\right>$ yield the frequency-dependent energies $\varepsilon^{\rm SF}_{\ell\boldsymbol{k}}(\omega)$, and  the SF reads
    \begin{equation}
    A(\omega) = \sum_{\ell\boldsymbol{k}}\delta(\omega-\varepsilon^{\rm SF}_{\ell\boldsymbol{k}}(\omega)).
    \label{eq:specfunc-1}
    \end{equation}
	
\begin{figure}[!tbp]
		\includegraphics[width=1.0\columnwidth]{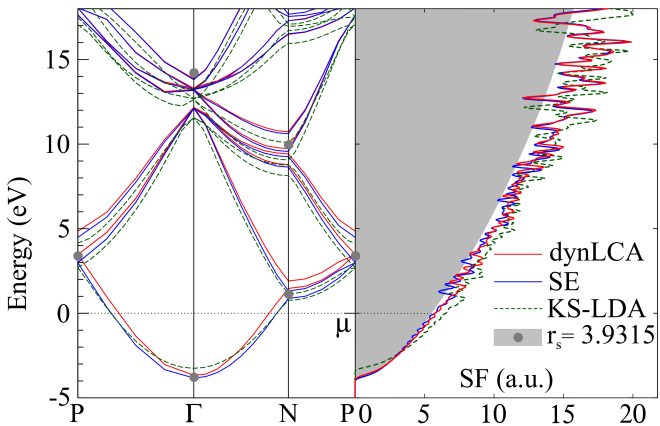}
		\caption{(Color online) Na band structure (left, in eV) and SF (right, atomic units). The blue curve is the HSE06 target result, while the red one is obtained with the dynamical local connector approximation Eq. \eqref{eq:pot+++}. The green curve is LDA--KS. For comparison, the HSE06 results in the HEG with $r_s=3.9315a_0$ are shown as  gray dots in the band structure and as shaded area for the SF. The zero of the energy scale is the Fermi energy.} 
		\label{fig:na_local}
	\end{figure}

\begin{figure}[!tbp]
		\includegraphics[width=1.0\columnwidth]{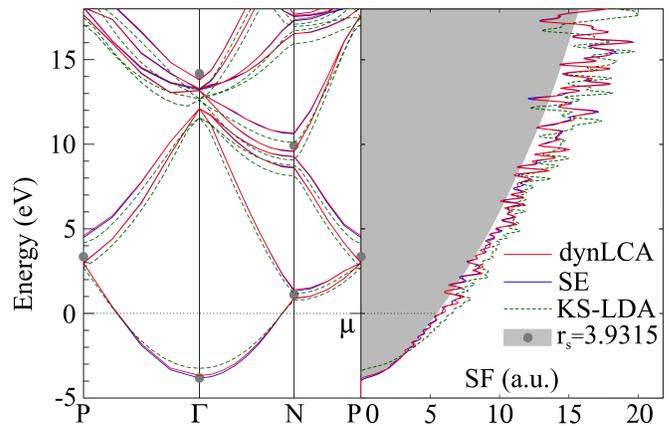}
		\caption{(Color online) Sodium as in Fig. \ref{fig:na_local}, but with the potential of Eq. \eqref{eq:QPcorr_pot} for the improved dynamical local connector approximation.}
		\label{fig:na_final}
	\end{figure}

\begin{figure}[!t]
		\includegraphics[width=1.0\columnwidth]{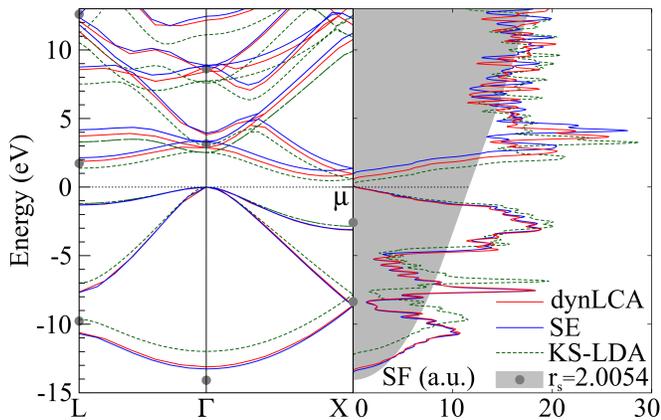}
		\caption{(Color online) Si band structure (left) and SF (right). Blue curve for the SE, red for dynLCA, Eq. \eqref{eq:pot+++} with Eq. \eqref{eq:QPcorr_pot}, green for LDA-KS; shaded area and gray dots for a SE calculation on a HEG of $r_s=2.0054a_0$.}
		\label{fig:si_final}
	\end{figure}

\begin{figure}[!t]
		\includegraphics[width=1.0\columnwidth]{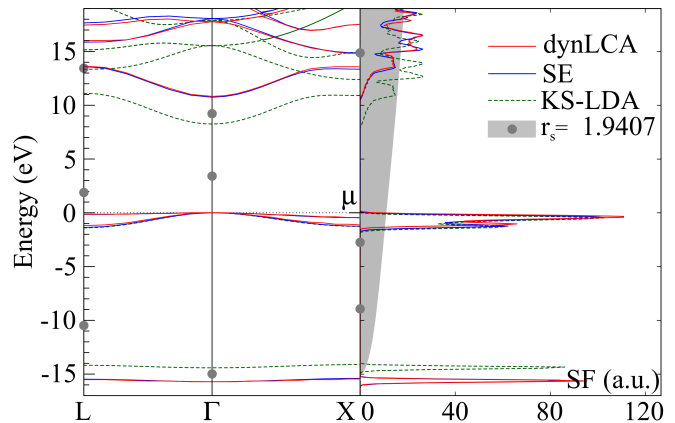}
		\caption{(Color online) Ar band structure (left) and SF (right). Blue curve for the SE, red for dynLCA, Eq. \eqref{eq:pot+++} with Eq. \eqref{eq:QPcorr_pot}, green for LDA-KS; shaded area and gray dots for a SE calculation on a HEG of $r_s=1.9407a_0$.}
		\label{fig:ar_final}
	\end{figure}

	To test this connector we compare the result of HSE06 calculations in real systems with results obtained for the same systems using the dynLCA $v_{\rm SF}$ (\ref{eq:pot+++}). Our test solids sodium, aluminum, silicon and solid argon range from simple metals to a covalent semiconductor and an insulator. We perform all calculations consistently in first-order perturbation theory on top of KS-LDA. The HSE06 calculations of the solids are our target results, given by the blue 
curves in figures
	\ref{fig:na_local},
	\ref{fig:na_final},
	\ref{fig:si_final} and
	\ref{fig:ar_final}.

Let us first look in the right panel of Fig. \ref{fig:na_local} at the SF of sodium, which is the material closest to the HEG. The non-local HSE06 self-energy yields a bandwidth of 3.83 eV, while an LDA calculation gives only 3.25 eV, which is off by 15\%. The dynLCA decreases the error to only  4\%, with a bandwidth of 3.66 eV. Also the overall agreement with the HSE06 result is quite good.
Only the small peaks in the SF at energies below 5 eV are blueshifted with respect to HSE06. 

By construction, the SP should reproduce the SF, but not necessarily the ${\boldsymbol{k}}$-resolved band structure (BS). 
Indeed, the dynLCA valence band in the left panel of Fig. \ref{fig:na_local} is clearly different from the HSE06 one, although the two SF are very similar. Still, the two BSs are 
surprisingly close. 
It is therefore interesting to further examine the link between the SF and the BS. In particular, the small peaks that are not in the correct position in dynLCA stem from  
lattice-related features in the BS, with gaps at high-symmetry ${\boldsymbol{k}}$-points (\emph{e.g.}, N and P, Fig. \ref{fig:na_local}) that are absent in the HEG. 
This rises the question whether one can further improve  the SF by improving the BS, while keeping the potential real and local. The answer is found in the BS of the HEG \cite{suppmat}, where one has a 	
one-to-one correspondence between $\varepsilon_{k}$ and $k$. In this condition, one can define a function $\Delta^h(\omega)$ such that a modified real spectral potential
	\begin{equation}
	v_{\rm SF}^h\left(\omega\right)\longrightarrow v_{SF}^h\left(\omega\right)-\Delta^h\left(\omega\right)
	\label{eq:QPcorr_pot}
	\end{equation}
	exactly reproduces the BS. 
	It is by construction the difference between the HEG SP and HSE06 BSs, namely $\Delta^h(\varepsilon^{\rm SF}_{k})\equiv\varepsilon^{\rm SF}_{k}-\varepsilon_k$. Now, what about the SF stemming from this modified SP \eqref{eq:QPcorr_pot}? The answer is subtle. Indeed, the new potential is different from the one defined in 
\cite{MatteoPRL}, and the SF calculated from  \eqref{eq:specfunc-1} would be wrong. However, since 
the BS is now exact by construction, 
one should use the \textit{energies} instead of the \textit{potential} to calculate the SF, \emph{i.e.}, one should use
\begin{equation}
A(\omega)=\sum_{\boldsymbol{k}}\delta\left(\omega-\varepsilon^{\rm SF}_{k}(\varepsilon^{\rm SF}_k)\right)
\label{eq:specfunc-2}
\end{equation}
    instead of \eqref{eq:specfunc-1}. With this, in the HEG the SF resulting from the SP is now again exactly the same as the HSE06 one \cite{suppmat}.
	%

For the real system, this leads to the following prescription: 
first, use the potential \eqref{eq:QPcorr_pot} in the dynLCA expression \eqref{eq:pot+++} (\emph{i.e.}, import also $\Delta^h$ from the HEG). For Na, this leads to the BS in the left panel of
Fig. \ref{fig:na_final}: it is in excellent agreement with the HSE06 reference. Second, we have to evaluate the spectral function using \eqref{eq:specfunc-2} instead of \eqref{eq:specfunc-1}. The right panel shows that the SF is now also extremely good, 
including the bandwidth and the position of the small peaks. 
Similar results are obtained for the less homogeneous metal Al \cite{suppmat}.

The big challenge, however, is to use the HEG-derived potential in order to describe very inhomogeneous, non-metallic systems. 
Indeed, whereas the connector might have been simplified for metals, its ingredients are essential for gapped systems, in particular, the DFT $v_{\rm xc}\left(\boldsymbol{r}\right)$ contribution that explicitly appears  in \eqref{eq:pot+++} and that is also found in the quasi-particle LDA approach to self-energy calculations \cite{Hedin1971,Wang,Pickett1984}.

	As a prototypical example, let us look at silicon: the LDA gap and valence bandwidth of respectively 0.56 eV and 11.96 eV are increased by the HSE06 calculation to the reference values of 1.20 eV and 13.26 eV. Our final dynLCA leads to the BS and SF in Fig. \ref{fig:si_final}. Both occupied and empty bands are significantly improved with respect to the LDA. In particular, the bandwidth error decreases from 10\% in the LDA to 1\% (13.11 eV) using dynLCA, and the gap error  from 53\% to 35\% (0.78 eV); also the shape of the SF is very good.
Note that  these encouraging results are obtained with a computational cost similar to that of the LDA calculation, and smaller than HSE06 by more than an order of magnitude. 
	
    Still, the silicon results are worse than those of the metals, and threaten a bad trend for even more inhomogeneous system with a larger gap. We therefore move to solid argon, an insulator where the LDA values for gap and bandwidth (8.31 eV and 14.41 eV, respectively) are increased to  10.74 eV and 15.71 eV, respectively, by the HSE06 (see Fig. \ref{fig:ar_final}). 
Surprisingly, dynLCA yields an almost perfect band structure and SF, with a gap of 10.85 eV and a bandwidth of 15.70 eV.
This reduces the gap error from 22\% in the LDA to 0.9\% in dynLCA, and the bandwidth error from 8.3\% in the LDA to 0.06\% in dynLCA. This inversion of the trend might be due to the fact that the metals are close to the HEG, while the electrons in Ar are quite localized and therefore more easily accessible by a connector based on the local density, while deviations only appear in an intermediate range represented by silicon.  

In conclusion, we have demonstrated that a local, real and frequency-dependent spectral potential (SP) can be used in practice to calculate the integrated spectral function (SF) with results similar to those of a non-local self-energy. Using the example of the HSE06 approximation, we have determined and tabulated the SP in the homogeneous electron gas. These freely available results can be used to build the SP in real materials according to an approximate, simple prescription, which we named dynamical local connector approximation (dynLCA). Our dynLCA calculations in several prototypical metals, semiconductors and insulators required a computational effort similar to that of the LDA, while leading to a significantly improved SF. Remaining discrepancies are mostly found in silicon, while results for the metals and the wide-gap insulator Ar are excellent. As a by-product, also the band structure is much better than the LDA one, which opens the possibility to describe even angle-resolved photoemission. 
	
	\begin{acknowledgments} 
		This research was supported by a Marie Curie FP7 Integration Grant within the 7th European Union Framework Programme and by the European  Research Council (ERC Grant Agreement n. 320971).
		Computational time was granted by GENCI (Project No. 544).
	\end{acknowledgments}


	\bibliographystyle{apsrev4-1}
	
	\bibliography{articlebib}
	
\end{document}